\documentstyle[prl,aps,psfig,multicol]{revtex}
\begin{document}
\draft
\title{Comment on ``Charged impurity scattering limited low temperature
resistivity of low density silicon inversion layers'' (Das~Sarma and
Hwang, cond-mat/9812216)}
\author{S.~V.~Kravchenko$^1$, D.~Simonian$^2$, and M.~P.~Sarachik$^3$}
\address{$^1$Physics Department, Northeastern University, Boston,
Massachusetts 02115}
\address{$^2$Department of Physics, Columbia University,
New York, New York 10027}
\address{$^3$Physics Department, City College of the City
University of New
York, New York, New York 10031}
\date{\today}
\maketitle
\begin{multicols}{2}
In a recent preprint, Das~Sarma and Hwang\cite{dassarma98} propose an
explanation for the sharp decrease in the $B=0$ resistivity at low
temperatures
which has been attributed to a transition to an unexpected
conducting phase in dilute high-mobility two-dimensional systems
(see Refs.[1-4] in \cite{dassarma98}).
The anomalous transport observed in these
experiments is ascribed in Ref.\cite{dassarma98} to
temperature-dependent
screening and energy averaging of the
scattering time.  The model yields curves that are
qualitatively similar to those observed experimentally: the resistivity
has a maximum at a temperature $\sim E_F/k_B$ and decreases at lower
temperatures by a factor of 3 to 10.  The
anomalous response to a magnetic field ({\it e.g.}, the increase
in low-temperature resistivity by
orders of magnitude~\cite{simonian97}), is not considered in
Ref.~\cite{dassarma98}.

Two main assumptions are made in the proposed model\cite{dassarma98}:
(1)~the transport behavior is dominated by charged impurity
scattering centers with a density $N_i$, and (2)~the metal-insulator
transition, which occurs
when the electron density ($n_s$) equals a critical density ($n_c$),
is due
to the freeze-out of $n_c$ carriers so that the net free carrier
density is given by $n\equiv n_s-n_c$ at $T=0$.  The
authors do not specify a mechanism for this carrier freeze-out and
simply accept it as an experimental fact.  Although not
included in their calculation, Das~Sarma and Hwang also note that their
model can be extended to include a thermally activated contribution
to the density of ``free'' electrons.

In this Comment, we examine whether the available experimental data
support the model of Das~Sarma and Hwang.

(i)~Comparison with the experimental data (see Fig.~1 of
Ref.~\cite{dassarma98}) is made
for an assumed density of charged impurities of
$3.5\times10^{9}$~cm$^{-2}$,
a value that is too small.  In an earlier publication, Klapwijk
and Das~Sarma~\cite{klapwijk98} explicitly stated that the number of
ionized impurities is ``$3\times10^{10}$~cm$^{-2}$ for high-mobility MOSFET's used for the 2D
MIT experiments.  There is very little room to vary this number by a
factor of two''.  Without reference to this earlier statement, the
authors now use a value for $N_i$ that is one order of magnitude
smaller~\cite{sample}.

(ii)~According to the proposed model, the number of ``free'' carriers
at zero temperature is zero at the ``critical'' carrier density
($n_s=n_c$) and it is very small ($n=n_s-n_c<<n_c$) near the transition. 
In this range, the transport must be dominated by thermally activated
carriers, which decrease exponentially in number as the temperature is
reduced.  It is known from experiment that at low temperatures the
resistance is independent of temperature\cite{krav,hanein98} at $n_c$
(the separatrix between the two phases) and depends weakly on
temperature
for nearby electron densities.  In order to give rise to a finite
conductivity $\sim e^2/h$ at the separatrix, an exponentially
small number of carriers must have an exponentially large mobility, a
circumstance that is rather improbable.

(iii)~Recent measurements of the Hall coefficient and Shubnikov-de~Haas
oscillations yield electron densities that are independent of
temperature and equal to $n_s$ rather than a density $n=n_s-n_c$ of ``free'' electrons~\cite{hanein98,hall}.  This implies that {\em all} the electrons contribute to the Hall conductance, including those that are frozen-out or localized.  Although this is known to occur in quantum systems such as Hall insulators, it is not clear why it can hold within the simple classical model proposed by Das~Sarma and Hwang.

\end{multicols}

\begin{thebibliography}{10}
\bibitem{dassarma98} S.~Das~Sarma and E.~H.~Hwang, preprint
cond-mat/9812216.
\bibitem{simonian97} D.~Simonian, S.~V.~Kravchenko, M.~P.~Sarachik, and
V.~M.~Pudalov, Phys.\ Rev.\ Lett.\ {\bf 79}, 2304 (1997); V.~M.~Pudalov,
G.~Brunthaler, A.~Prinz, and G.~Bauer, JETP Lett.\ {\bf 65},  932
(1997).
\bibitem{klapwijk98} T.~M.~Klapwijk and S.~Das~Sarma, preprint
cond-mat/9810349.
\bibitem{sample} We note that although sample Si-15 had a particularly
high peak mobility at 4.2~K (almost twice that of other samples), this cannot account for a reduction in $N_i$ by a factor of 10.  The density
of charged traps $N_i$ in samples of comparable quality was
estimated to be $1.5\times10^{10}$~\cite{pudalov93}.
\bibitem{pudalov93} V.~M.~Pudalov, M.~D'Iorio, S.~V.~Kravchenko, and
J.~W.~Campbell, Phys.\ Rev.\ Lett.\ {\bf 70}, 1866 (1993).
\bibitem{krav} S.~V.~Kravchenko, W.~E.~Mason, G.~E.~Bowker,
J.~E.~Furneaux, V.~M.~Pudalov, and M.~D'Iorio, Phys.\ Rev.\ B {\bf 51},
7038 (1995).
\bibitem{hanein98} Y.~Hanein, D.~Shahar, J.~Yoon, C.~C.~Li, D.~C.~Tsui,
and H.~Shtrikman, Phys.\ Rev.\ B {\bf 58}, R7520 (1998).
\bibitem{hall} D.~Simonian, K.~Mertes, M.~P.~Sarachik, S.~V.~Kravchenko,
and T.~M.~Klapwijk, in preparation.
\end{thebibliography}
\end{document}